\documentclass[10pt,twocolumn,letterpaper]{article}

\usepackage{amsthm, amssymb}
\usepackage{multirow}

\usepackage[pagenumbers]{iccv} % To force page numbers, e.g. for an arXiv version

\definecolor{iccvblue}{rgb}{0.21,0.49,0.74}
\usepackage[pagebackref,breaklinks,colorlinks,allcolors=iccvblue]{hyperref}

\def\paperID{11646} % *** Enter the Paper ID here
\def\confName{ICCV}
\def\confYear{2025}

\title{
AFUNet: Cross-Iterative Alignment-Fusion Synergy for HDR Reconstruction via Deep Unfolding Paradigm
}

\author{
Xinyue Li$^1$,
Zhangkai Ni$^{1\ast}$, 
Wenhan Yang$^2$ \\
$^1$Tongji University, 
$^2$Pengcheng Laboratory \\
{\tt\small 2252065@tongji.edu.cn, zkni@tongji.edu.cn, yangwh@pcl.ac.cn}
}

\begin{document}
\maketitle

\let\thefootnote\relax\footnotetext{\noindent$^\ast$Corresponding author.}

\begin{abstract}

Existing learning-based methods effectively reconstruct HDR images from multi-exposure LDR inputs with extended dynamic range and improved detail, but often rely on empirical design rather than a theoretical foundation, which can impact their reliability. 
To address these limitations, we propose the cross-iterative Alignment and Fusion deep Unfolding Network (AFUNet), where HDR reconstruction is systematically decoupled into two interleaved subtasks—alignment and fusion—optimized through alternating refinement, achieving synergy between the two subtasks to enhance the overall performance.
Our method formulates multi-exposure HDR reconstruction from a Maximum A Posteriori (MAP) estimation perspective, explicitly incorporating spatial correspondence priors across LDR images and naturally bridging the alignment and fusion subproblems through joint constraints.
Building on the mathematical foundation, we reimagine traditional iterative optimization through unfolding—transforming the conventional solution process into an end-to-end trainable AFUNet with carefully designed modules that work progressively.
Specifically, each iteration of AFUNet incorporates an Alignment-Fusion Module (AFM) that alternates between a Spatial Alignment Module (SAM) for alignment and a Channel Fusion Module (CFM) for adaptive feature fusion, progressively bridging misaligned content and exposure discrepancies.
Extensive qualitative and quantitative evaluations demonstrate AFUNet’s superior performance, consistently surpassing state-of-the-art methods.
Our code is available at: \url{https://github.com/eezkni/AFUNet}

\end{abstract}

\section{Introduction}
\label{sec:intro}

\begin{figure}[t]
\centering
\includegraphics[width=1.0\linewidth]{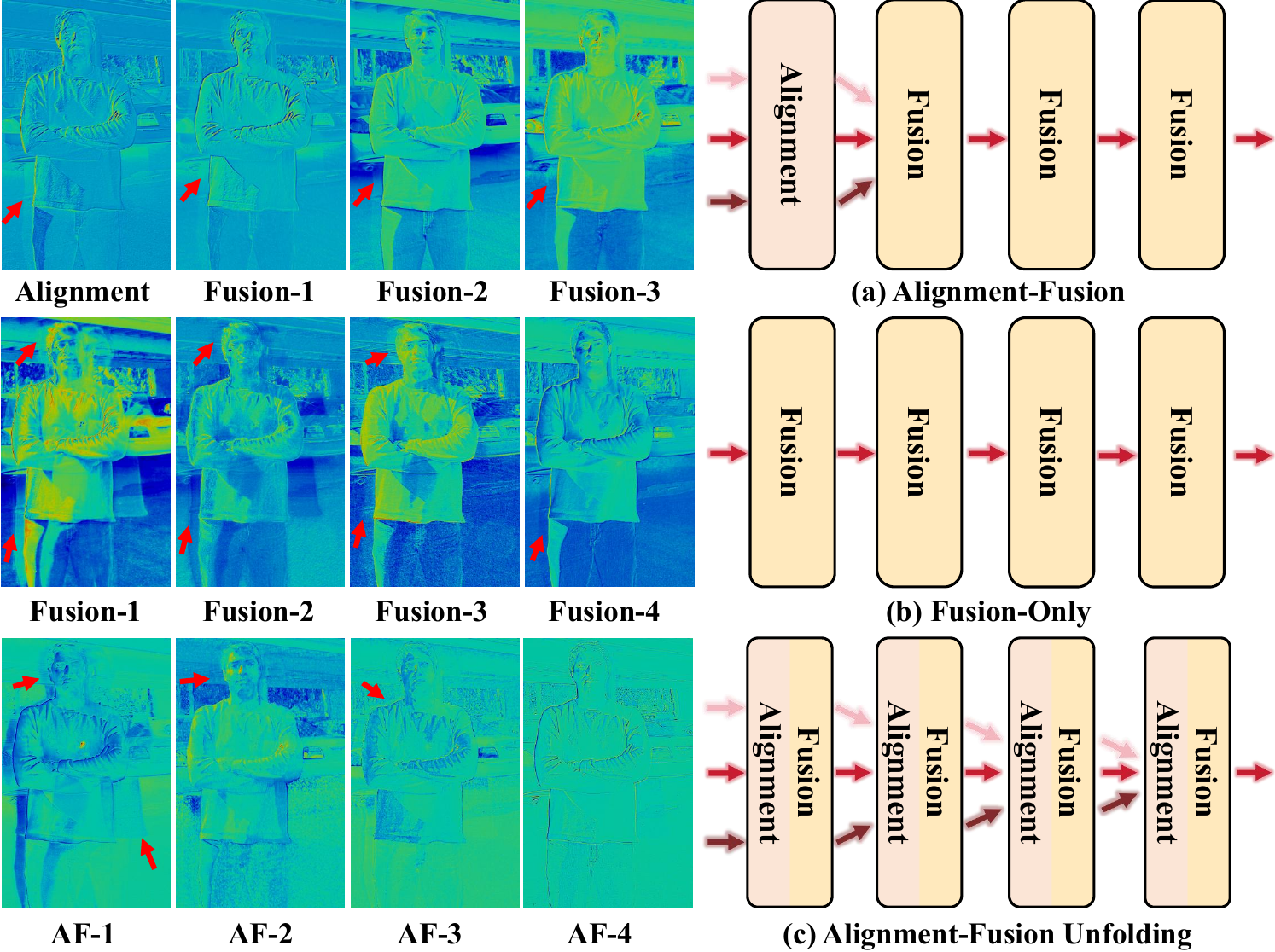}
\vspace{-15pt}
\caption{
Comparison of three HDR reconstruction paradigms: 
(a) the ``Alignment-Fusion" paradigm, (b) the ``Fusion-Only" paradigm, and (c) our proposed ``Alignment and Fusion Unfolding" paradigm. 
The feature maps from the reconstruction process of the three paradigms are shown on the left. (c) shows superior deghosting due to iterative alignment that continuously corrects misalignment, while (a) leaves artifacts due to pre-alignment, and (b) lacks explicit alignment, resulting in less effective deghosting.
}
\label{fig:teaser}
\vspace{-10pt}
\end{figure}

Multi-exposure High Dynamic Range (HDR) imaging aims to effectively leverage information from multiple Low Dynamic Range (LDR) images captured at varying exposures to reconstruct a larger dynamic range HDR image~\cite{debevec1997recovering}. 
HDR images possess a broader dynamic range, offering a realistic and visually appealing experience, making it indispensable for applications such as satellite remote sensing, virtual reality and autonomous driving.

While existing HDR methods can produce accurate results when LDR images are well-aligned~\cite{debevec1997recovering, zhang2011gradient, Prabhakar2017DeepFuse, Ma2020Deep}, misalignment issues caused by dynamic scenes or camera jitter are prone to induce ghosting artifacts. 
Traditional methods make effort in either rejecting misaligned pixels~\cite{grosch2006fast, zhang2011gradient, oh2014robust}, merging images at the patch level~\cite{sen2012robust, hu2013hdr, ma2017robust, lee2014ghost}, or performing explicit alignment of LDR images~\cite{bogoni2000extending, kang2003high, zimmer2011freehand}. 
However, these methods are still far from being practical. 
Rejection-based methods might miss important details in moving regions, patch-based methods are computationally intensive, while alignment-based methods heavily depend on precise alignment, which is challenging under large motion or poor exposure conditions.

With the rise of deep learning-based methods, leveraging data priors and flexible modeling has allowed more effective solutions to these challenges.
They generally follow two main paradigms, as shown in Fig.~\ref{fig:teaser}. 
The majority of methods~\cite{liu2021adnet, yan2019attention, yan2023unified,  chen2023improving, zhang2024hl} adopt a two-stage paradigm: first utilizing an alignment module to align LDR images and then refine through a complex network to fuse features and reconstruct the HDR image, which is referred to as the ``Alignment-Fusion” paradigm. 
In contrast, the other branch~\cite{ tel2023alignment, yan2020deep, ye2021progressive} bypasses alignment, instead directly fusing features implicitly through stacked modules, summarized as the ``Fusion-Only” paradigm. 
While both paradigms have proven effective, the former often loses information during the alignment process, while the latter lacks explicit alignment, leading to ghosting. Additionally, both paradigms are designed empirically and lack a mathematical foundation.
To address these issues, we propose a novel approach that performs alignment and fusion alternately. 
We formulate HDR imaging from the MAP view and translate this formulation into an end-to-end trainable network through unfolding the iterative optimization steps.

In this study, we propose a cross-iterative alignment and fusion deep unfolding network, achieving superior performance. 
Our key insight is to decouple the complex HDR reconstruction process into alignment and fusion subproblems with two prior regularization terms to capture the spatial correspondence among multi-exposure LDR images.
Then we turn the iterative reconstruction steps into a fixed number of designed concatenated blocks. 
Unlike prior deep unfolding methods that take HDR imaging as a low-rank completion problem~\cite{mai2022deep}, our method is more flexible and relaxed, accommodating more real-world scenarios without relying on additional or strict assumptions. 
The main contributions are as follows:
\begin{itemize}
    \item
    \textbf{A Unified MAP View for HDR Reconstruction}: We formulate HDR reconstruction as a Maximum A Posteriori (MAP) estimation problem, introducing prior regularization terms that constrain the spatial correspondence among LDRs to decompose the task into alignment and fusion subproblems, which are solved in a cross-iterative manner for achieving high-quality results.
    \item 
    \textbf{End-to-End Theory-Grounded Unfolding Network}: We develop a cross-iterative Alignment and Fusion deep Unfolding Network (\textbf{AFUNet}) that allows end-to-end training for HDR reconstruction. It comprises stacked Alignment Fusion Modules (AFMs), each corresponding to an iterative step derived from mathematical modeling.
    \item 
    \textbf{State-of-the-Art Performance}: Extensive qualitative and quantitative experiments demonstrate that AFUNet achieves state-of-the-art performance in HDR reconstruction, delivering visually appealing results that align with human perceptual aesthetics.
    
\end{itemize}

\section{Related Work}
\label{sec:Related Work}

\subsection{Learning-based HDR Reconstruction Methods}

Most existing deep learning-based HDR reconstruction methods follow the ``Alignment-Fusion" paradigm. 
Early non-end-to-end methods~\cite{kalantari2017deep} first align LDR images using optical flow before fusing them to produce HDR images. Although Kalantari \textit{et al.}~\cite{kalantari2017deep} shows notable improvements over traditional approaches, it remains error-prone in handling complex foreground motions. 
Subsequent techniques improve alignment through advanced motion estimation modules~\cite{catley2022flexhdr, kong2024safnet} or by employing attention-based feature alignment within Convolutional Neural Network (CNN) frameworks~\cite{liu2021adnet, yan2019attention}. 
More recently, transformer-based methods~\cite{liu2022ghost, yan2023unified, song2022selective} leveraged multi-head self-attention mechanisms for enhanced alignment and fusion capabilities. Additionally, some methods~\cite{ tel2023alignment, yan2020deep, ye2021progressive} bypass the alignment, performing only feature fusion. 
While these paradigms achieve promising performance, their deep neural network architectures are often empirically designed. 
Deep Unfolding Networks (DUNs) have gained traction for HDR tasks, balancing the advantages of model-based and learning-based approaches while addressing their limitations. 
In this work, we introduce a novel deep unfolding paradigm that combines the strengths of the ``Alignment-Fusion" paradigm with model-based techniques, resulting in a cross-iterative alignment and fusion network.
Unlike MERF~\cite{hong2024merf}, which uses a two-stage approach with separate pre-training for alignment and fusion, followed by Generative Adversarial Network (GAN)-like iterative training, our method is a single-stage end-to-end training process.

\subsection{Deep Unfolding Networks}

DUNs have shown strong performance on inverse problems like Super-Resolution~\cite{zhang2020deep, huang2020unfolding, zhou2023memory}, Compressive Sensing~\cite{wang2024progressive, song2023dynamic, guo2024cpp}, and Pan-Sharpening~\cite{wang2024deep, yang2022memory, meng2024progressive} by unfolding model-based iterative algorithms into end-to-end optimized deep networks. 
However, DUNs are relatively underexplored in the multi-exposure HDR imaging field.
Mai \textit{et al.}~\cite{mai2022deep} applied an unrolling strategy for HDR imaging, modeling it through low-rank tensor completion to construct an interpretable deep network. 
Though Mai \textit{et al.}~\cite{mai2022deep} is effective in some cases, it is limited in HDR reconstruction because its low-rank modeling oversimplifies complex scenes, misses crucial details, and does not fully utilize available information.
In contrast, our method proposes a simpler but effective model that fully leverages multi-exposure information and provides a more general solution for multi-exposure HDR imaging. 
Our carefully designed deep unfolding approach achieves both qualitative and quantitative improvements, delivering state-of-the-art HDR reconstruction results.

\section{Methodology}

\subsection{Motivation}

Given a set of LDR multi-exposure images, we aim to address the following two issues:
\begin{itemize}
    \item \textbf{Limited Deghosting Effectiveness.} The “Alignment-Fusion” paradigm, based on pre-alignment, struggles with LDRs due to motion-induced misalignments and information loss in over- or under-exposed regions, which hinder precise alignment. While pre-alignment has shown some efficacy, relying solely on this strategy is less than ideal for HDR reconstruction. Integrating alignment directly into the fusion process holds the promise of more effectively suppressing ghosting artifacts.
    \item \textbf{Lack of Mathematical Foundation.} Deep learning has propelled significant advances in HDR reconstruction, but various prevailing architectures lack mathematical foundation, often being empirically constructed. 
\end{itemize}

\noindent Our core goal is to develop a well-structured model specifically tailored for HDR reconstruction that effectively aligns and fuses multi-exposure images progressively to produce high-quality HDR outputs. 
In the following sections, we introduce our proposed method in detail.

\subsection{Problem Formulation}
\label{sec:ProblemFormulation}

The degradation process from an HDR image to an LDR image can be mathematically formulated as $y = Dx+n$, where $x$ denotes the HDR image, $D$ represents the degradation transformation, $y$ is the LDR image, and $n$ is the additive noise. 
The intractable ill-posed problem of reconstructing $x$ is reformulated as an optimization problem under the MAP framework, including data fidelity and regularization terms $\rm \Psi(\cdot)$. 
The data fidelity term is typically defined as the \( \ell_2 \) norm, expressed in the following energy function:
\begin{equation} 
\small
\hat{x} = \underset{x}{\arg\min} \frac{1}{2} \|y - Dx\|^2_2 + {\rm{\lambda}} {\rm{\Psi}}(x),
\label{eq:energyFunc}
\end{equation}
where $\hat{x}$ is the reconstructed HDR image, $\rm{\lambda}$ is a regularization weighting hyperparameter. 
We aim to merge three differently exposed LDR images (\textit{i.e.}, under-exposed image $y_1$, normal-exposed image $y_2$, and over-exposed image $y_3$) into a single high-quality HDR image $\hat{x}$ without artifacts. 
Specifically, the LDR image $y_2$ serves as the reference image, and the predicted HDR image must be content-aligned with $y_2$. 
Accordingly, we extend Eq.~\eqref{eq:energyFunc} by introducing the non-reference LDR images $y_1$ and $y_3$, which provide complementary scene details from their distinct exposure levels and thereby enhance the reconstruction of $x$ from $y_2$. 

However, directly applying the priors from the non-reference LDR images might limit their effective utilization due to potential misalignment between the LDR images. 
To address this issue, we introduce two spatial correspondence prior regularization terms, ${\rm{p}}_1 \left( y_2, {\rm \alpha}_{1} \right)$ and ${\rm{p}}_3 \left( y_2, {\rm \alpha}_{3} \right)$, which explicitly model the rich priors between the non-reference LDR images $y_1, y_3$ and the reference LDR image $y_2$ for HDR reconstruction. 
The $\alpha_1$ and $\alpha_3$  represent the spatially aligned versions of $y_1$ and $y_3$, respectively, which are iteratively optimized to align the structure and content with $x$.
Thus, we reformulate the optimization problem for HDR reconstruction as follows:
\begin{equation}
\small
\underset{x,\alpha_1,\alpha_3}{\arg\min}
\|y_2 - D_2x\|^2_2 
+ \lambda_1 {\rm{p}}_1 \left( D_1x, \alpha_1 \right) 
+ \lambda_3 {\rm{p}}_3 \left( D_3x, \alpha_3 \right),
\label{eq:energyFuncFinal}
\end{equation}
where $\lambda_1, \lambda_3$ are balancing coefficients, and $D_1,D_3$ are the degradation transformations of $y_1,y_3$, respectively.

To solve this model efficiently, we first decompose Eq.~\eqref{eq:energyFuncFinal} into two subproblems—alignment and fusion—and solve them alternately:
\begin{subequations}
\small
\begin{align}
\alpha_1^t &= \underset{\alpha_1}{\arg\min} \,
{\rm{p}_1} (D_1x^{t-1}, \alpha_1), 
\label{eq:sub-problem-a}\\
\alpha_3^t &= \underset{\alpha_3}{\arg\min} \,
{\rm{p}_3} (D_3x^{t-1}, \alpha_3), 
\label{eq:sub-problem-b} \\ 
x^t = & \underset{x}{\arg\min} 
\frac{1}{2} \|y_2 - D_2x\|^2_2  + \lambda_1 {\rm{p}_1} (D_1 x, \alpha_1^t) + \lambda_3 {\rm{p}_3} (D_3x, \alpha_3^t).
\label{eq:sub-problem-c} 
\end{align}
\end{subequations}
\vspace{-10pt}

\begin{figure*}[t]
\hsize=\textwidth
\centering
\includegraphics[width=1.0\textwidth]{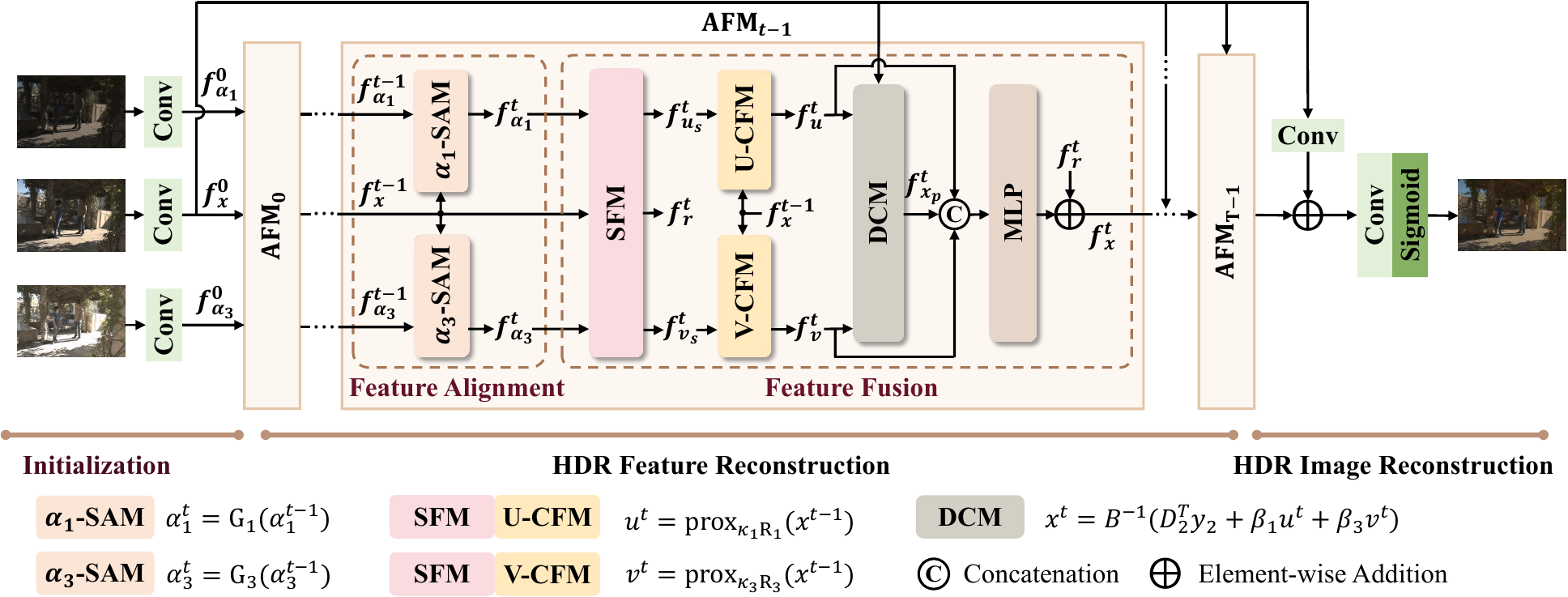}
\vspace{-15pt}
\caption{
Framework of the AFUNet for HDR reconstruction consists of three main processes: Initialization, HDR Feature Reconstruction, and HDR Image Reconstruction. 
Within HDR Feature Reconstruction, stacked Alignment Fusion Modules (AFMs) iteratively refine target HDR features via alternating alignment and fusion.
In the Feature Alignment subprocess, $f_{\alpha_1}^{t-1}$ and $f_{\alpha_3}^{t-1}$ are aligned with $f_x^{t-1}$ by the Spatial Alignment Modules—denoted as $\alpha_1$-SAM and $\alpha_3$-SAM. 
In the Feature Fusion subprocess, the Spatial Fusion Module (SFM) performs preliminary optimization to obtain $f_{u_s}^t$ and $f_{v_s}^t$, followed by the Channel Fusion Modules—denoted as U-CFM and V-CFM—obtain $f_{u}^t$ and $f_{v}^t$, respectively. 
Finally, the Data Consistency Module (DCM) obtains $f_{x_p}^t$, and an MLP with residual addition further refines $f_x^t$ before it is transmitted to the next stage. 
The $\kappa_i= \frac{\lambda_i}{\beta_i}$, ${\rm R}_i={{\rm p}_i}(\cdot,\alpha_i^t)$ ($i=1,3$), and $B^{-1}=(D^{\rm T}_2D_2+(\beta_1+\beta_3)I)^{-1}$.
}
\vspace{-10pt}
\label{fig:model}
\end{figure*}

For the alignment subproblem in Eq.~\eqref{eq:sub-problem-a} and Eq.~\eqref{eq:sub-problem-b}, we define gradient descent operators ${\rm G}_i(\cdot) = \alpha_i^{t-1} - \varsigma_i	\triangledown_{\alpha_i} {\rm p_i} (D_i x^{t-1}, \alpha_i^{t-1})$, where $\varsigma_i$ is the step size,  $\triangledown_{\alpha_i} {\rm p_i} (D_i x^{t-1}, \alpha_i^{t-1})$  are the gradient of the spatial correspondence prior term for the aligned variables $\alpha_i$, $i=1,3$. The $t$-th optimization step can be expressed as:
\begin{subequations}
\small
\begin{align}
&\alpha_1^t = {\rm G}_1(\alpha_1^{t-1}),
\label{eq:alignment-gd-a}  \\
&\alpha_3^t ={\rm G}_3(\alpha_3^{t-1}).
\label{eq:alignment-gd-b}
\end{align}
\end{subequations}

For the fusion subproblem in Eq.~\eqref{eq:sub-problem-c}, we solve it using the Half Quadratic Splitting (HQS)~\cite{geman1995nonlinear} method to decouple the data fidelity term and regularization terms. 
First, we introduce two auxiliary variables, $u$ and $v$, corresponding to the two prior regularization terms that constrain the spatial correspondence among different LDR images. 
We impose constraints to ensure that $u$ and $v$ are as close as possible to the target image $x$:
\begin{equation} 
\small
\begin{aligned}
&\underset{x,u,v}{\arg\min} 
\frac{1}{2}\|y_2 - D_2x\|^2_2 
+ 
\lambda_1 {\rm{p}}_1 \left( D_1u, \alpha_1^t \right) \\
& + 
\lambda_3 {\rm{p}}_3 \left( D_3v, \alpha_3^t \right) 
+ 
\frac{\beta_1}{2} \| u-x \|^2_2
+
\frac{\beta_3}{2} \| v-x \|^2_2,
\end{aligned}
\label{eq:hqs-fusion-sub-problem}
\end{equation}
where $\beta_1$ and $\beta_3$ are weighting factors for the added terms.

The fusion problem can then be split into three subproblems, which are updated iteratively:
\begin{subequations}
\footnotesize
\begin{align}
& u^{t} =  \underset{u}{\arg\min}\frac{\beta_1}{2} \| u- x^{t-1} \|^2_2 + \lambda_1 {\rm{p}}_1 (D_1 u,{\alpha_1^t}),
\label{eq:hqs-a}\\
&v^{t} =  \underset{v}{\arg\min}\frac{\beta_3}{2}\| v- x^{t-1} \|^2_2 + \lambda_3 {\rm{p}}_3 (D_3 v, {\alpha_3^t}), \label{eq:hqs-b}\\
& x^{t} = \underset{x}{\arg\min} \frac{1}{2} \|y_2 - D_2x\|^2_2 + \frac{\beta_1}{2} \| u^{t} - x\|^2_2 + \frac{\beta_3}{2} \| v^{t} - x\|^2_2 .
\label{eq:hqs-c} 
\end{align}
\end{subequations}

Given reconstruction target image $x^{t-1}$ and aligned image $\alpha_1^t$ and $\alpha_3^t$, we define proximal operators ${\rm prox}_ {\frac{\lambda_1}{\beta_1} {\rm p_1}(\cdot)}(\cdot)$ and ${\rm prox}_ {\frac{\lambda_3}{\beta_3} {\rm p_3}(\cdot)}(\cdot)$ for the optimization of $u^t$ and $v^t$: 
${\rm{prox}}_{\frac{\lambda_1}{\beta_1} {\rm{p_1}} \left( 
 \cdot,\alpha_1^t) \right) } (x^{t-1})= \underset{u}{\arg\min}\frac{\beta_1}{2} \| u-x^{t-1} \|^2_2+\lambda_1 {\rm{p}}_1 (D_1 u,{\alpha_1^t}) $ and ${\rm{prox}}_{\frac{\lambda_3}{\beta_3} {\rm{p_3}} \left( 
 \cdot,\alpha_3^t) \right) } (x^{t-1}) = \underset{v}{\arg\min}\frac{\beta_3}{2} \| v- x^{t-1} \|^2_2 + \lambda_3 {\rm{p}}_3 (D_3 v,{\alpha_3^t})$. Eq.~\eqref{eq:hqs-a} and Eq.~\eqref{eq:hqs-b} is then solved by the following equation:
\begin{subequations}
\small
\begin{align}
& u^{t} = {\rm{prox}}_{\frac{\lambda_1}{\beta_1} {\rm{p_1}} \left( 
 \cdot,\alpha_1^t) \right) } (x^{t-1}),
\label{eq:pgd-a}\\
&v^{t} = {\rm{prox}}_{\frac{\lambda_3}{\beta_3} {{\rm{p}}_3} \left( 
 \cdot,\alpha_3^t) \right) } (x^{t-1}). 
\label{eq:pgd-b}
\end{align}
\end{subequations}

Eq.~\eqref{eq:hqs-c} represents a quadratic regularized least squares problem, which has a closed-form solution:
\begin{equation}
\small
\begin{aligned}
x^{t} = (D_2^T D_2 + (\beta_1 + \beta_3) I)^{-1} (D_2^T y_2 + \beta_1 u^{t} + \beta_3 v^{t} ),
\end{aligned}
\label{eq:hqs-update-X-M}
\end{equation}
where $I$ is the identity matrix, $D_2^T$ designates the transposition of degradation transformation matrix $D_2$. 
The matrix inverse is computationally expensive, so we treat $(D_2^T D_2 + (\beta_1 + \beta_3) I)^{-1}$ as a single entity, denoted as $B^{-1}$. 
To efficiently handle this, we design neural networks to learn the complex degradation matrices $B^{-1}$ and $D_2^T$.

\subsection{Deep Unfolding Network}

The AFUNet pipeline, as illustrated in Fig.~\ref{fig:model}, consists of three primary processes: \textbf{Initialization}, \textbf{HDR Feature Reconstruction}, and \textbf{HDR Image Reconstruction}. 
The feature reconstruction process includes $\rm T$ stages for optimization. Each stage can be further subdivided into two key subprocesses: \textbf{Feature Alignment} and \textbf{Feature Fusion}.
Notably, as outlined in Section~\ref{sec:ProblemFormulation}, the solution to the unfolding paradigm is performed in the image space, where the LDR images $y_1, y_2, y_3$ are directly involved in the optimization process. 
In contrast, in this section, we apply the iterative optimization and refinement at the feature space and propose a learnable solution using a deep unfolding network. \textit{The algorithm of AFUNet is available in Section 2 of the supplementary material.} The details are as follows:

\begin{figure*}[t]
\hsize=\textwidth
\centering
\includegraphics[width=1.0\textwidth]{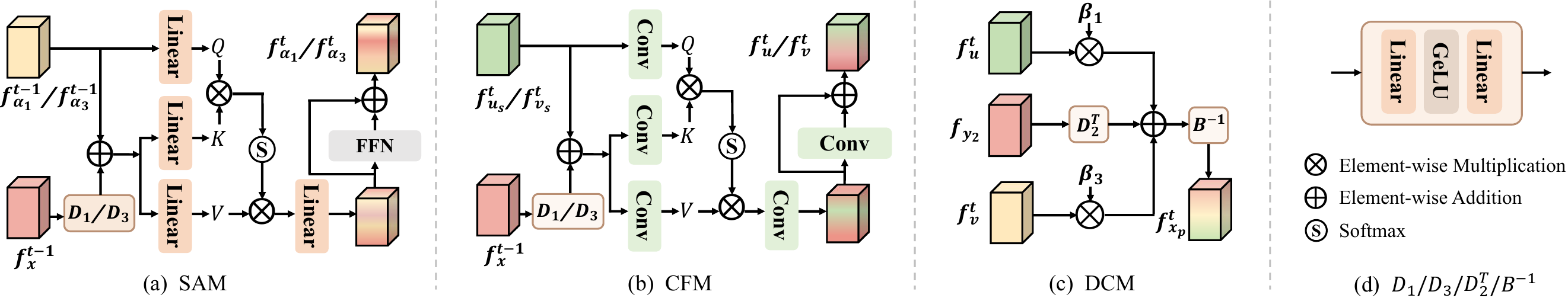}
\vspace{-16pt}
\caption{
(a) SAM uses window-based cross-attention to align $f_{\alpha_1}^{t-1}$ or $f_{\alpha_3}^{t-1}$ with $f_x^t$, to obtain $f_{\alpha_1}^{t}$ or $f_{\alpha_3}^t$. 
(b) CFM fuses $f_{u_s}^{t}$ or $f_{v_s}^{t}$ with $f_x^{t-1}$, to obtain $f_u^t$ or $f_v^t$. 
(c) DCM updates $f_x^t$ to obtain $f_{x_p}^t$ using $f_u^t, f_v^t$.
(d) The degradation transformations, including $D_1$, $D_3$, $D_2^T$, and $B^{-1}$, are learned using independent MLPs. 
}
\vspace{-12pt}
\label{fig:moduleA}
\end{figure*}

\textbf{1) Initialization.} 
Given the input images ${y_i} = [L_i, H_i] \in {\mathbb{R}}^{B\times 6\times H\times W} (i = 1,2,3)$, where $L_i$ is the LDR image and $H_i$ refers to the gamma-corrected result of $L_i$ that provides additional information, these images are projected into the feature domain $f_{y_i} \in {\mathbb{R}}^{B\times C\times H\times W} (i = 1,2,3)$ through three shallow feature extraction modules ${\rm SFEM}_i(\cdot)$, respectively. 
$B$, $C$, $H$, and $W$ denote the batch size, number of channels, height, and width of the feature maps, respectively. 
The feature extraction is expressed as:
\begin{equation} 
f_{{y_i}} = {{\rm SFEM}_i}({y_i}), \quad i=1,2,3,
\label{eq:shallowFeatureExtraction}
\end{equation}
where \({\rm SFEM}_i(\cdot)\) is a convolutional layer with a \(3 \times 3\) kernel.
The feature maps $f_{{y_1}}, f_{{y_2}}, f_{{y_3}}$ are used to initialize the features $f_{\alpha_1}^0, f_x^0, f_{\alpha_3}^0$, respectively, which serve as the initial input features for the HDR feature reconstruction process.

\vspace{1mm}
\textbf{2) HDR Feature Reconstruction.} 
We propose $\rm T$ stages using the stacked Alignment Fusion Modules (AFMs), which are unfolded from our iterative optimization algorithms to solve the HDR reconstruction objective. 
Importantly, all stages share the same structure but have independent parameters. 
The iterative process is expressed as:
\begin{equation} 
\small
f_{\alpha_1}^t, f_{x}^{t}, f_{\alpha_3}^t = {\rm{AFM}}_{t-1}(f_{\alpha_1}^{t-1},f_{x}^{t-1},f_{\alpha_3}^{t-1}), \\
\label{eq:AFM}
\end{equation}
where $t=1,2,\cdots, \rm T$. The features $f_{\alpha_1}^{t-1}, f_{x}^{t-1}, f_{\alpha_3}^{t-1}$ represent the outputs from stage $t-2$, \textit{i.e.}, the previous stage, and $f_{\alpha_1}^{t}, f_{x}^{t}, f_{\alpha_3}^{t}$ are the outputs from stage $t-1$.

\vspace{1mm}
\textbf{A. Feature Alignment Subproblem.} 
We design a simple but effective Spatial Alignment Module (SAM) to align $f_{\alpha_1}^{t-1}, f_{\alpha_3}^{t-1}$ with the intermediate reconstructed feature $f_{x}^{t-1}$. There are two SAMs act as gradient descent operators $\rm G1(\cdot)$ and $\rm G_3(\cdot)$ in Eq.~\eqref{eq:alignment-gd-a} and ~\eqref{eq:alignment-gd-b}, respectively, producing the aligned features $f_{\alpha_1}^t,f_{\alpha_3}^t$.
As shown in Fig.~\ref{fig:moduleA} (a), we construct the SAM as a window-based cross-attention transformer block~\cite{liu2021swin}, which can be formulated as: 
\begin{equation} 
\small
f_{\alpha_i}^t = {\rm FFN} ({\rm{WCAA}}(f_{\alpha_i}^{t-1},f_{x}^{t-1})), 
\label{eq:transformer-cross-attention}
\end{equation}
where $\rm{WCAA} (\cdot)$ is the Window-based Cross-Attention Alignment module, and $\rm{FFN}(\cdot)$ is a Feed-Forward Network, with \(i = 1, 3\). 
This enables us to query the spatial information in the reference image feature and use $f_{x}^{t-1}$ to preserve the spatial structure of the reference image.
\begin{equation} 
\small
{\rm{WCAA}}(f_{\alpha_i}^{t-1},f_{x}^{t-1}) = {\rm{Softmax}}(\frac{Q K^{T}}{\sqrt{{d}_{k}}})V,
\label{eq:transformer-cross-attention}
\end{equation}
where
\begin{equation*}
\small
\begin{aligned}
& {Q}={\rm W}_{Q} (f_{\alpha_i}^{t-1}), \\
& {K}={\rm W}_{K} (f_{\alpha_i}^{t-1}+{{\rm MLP}_{D_i}}(f_x^{t-1})), \\
& {V}={\rm W}_{V} (f_{\alpha_i}^{t-1}+{{\rm MLP}_{D_i}}(f_x^{t-1})),
\end{aligned}    
\end{equation*} 
${\rm W}_{Q} (\cdot)$,${\rm W}_{K} (\cdot)$ and ${\rm W}_{V} (\cdot)$ are learnable transformations, ${d}_{k}$ is the feature channel dimension of ${Q}$ and ${K}$, ${\rm MLP}_{D_i}(\cdot)$ denotes ${\rm MLP}(\cdot)$ for learning degradation transformations $D_i$ ($i=1,3$) as shown in Fig.~\ref{fig:moduleA} (d).
Notably, we use window-based cross-attention to focus on local spatial information between features, as alignment primarily targets high-frequency and structural details. In contrast, global transformations, such as warping, lead to suboptimal alignment results due to their application of uniform transformations across the entire image.

\vspace{1mm}
\textbf{B. Feature Fusion Subproblem.} 
It is involves the updating of three variables: $f_{u}^t$, $f_{v}^t$, and $f_{x}^t$.

\textbf{Update $f_{u}^t$ and $f_{v}^t$}: 
For the optimization of $f_{u}^t$ and $f_{v}^t$ according to Eq.~\eqref{eq:pgd-a} and Eq.~\eqref{eq:pgd-b}, we propose a two-step fusion module as the proximal operator for learning priors between features. First, it fuses features spatially and then conducts channel-wise fusion to integrate $f_{x}^{t-1}$ with $f_{\alpha_1}^{t}$ and $f_{\alpha_3}^{t}$.
This process can be expressed as:
\begin{subequations}
\small
\begin{align}
f_{u_s}^t, f_r^t, f_{v_s}^t  &= {\rm SFM}(f_{\alpha_1}^{t},f_{x}^{t-1},f_{\alpha_3}^{t}), 
\label{eq:spatial-fusion} \\
f_{u}^t &= {\rm U}\text{-}{\rm CFM}(f_{u_s}^t, f_{x}^{t-1}), 
\label{eq:channel-fusion-u} \\
f_{v}^t &= {\rm V}\text{-}{\rm CFM}(f_{v_s}^t, f_{x}^{t-1}).
\label{eq:channel-fusion-v}
\end{align}
\end{subequations}

\textit{1) Spatial Fusion Module.} The $\rm{SFM} (\cdot)$ is a Transformer-based Spatial Fusion Module where $f_{\alpha_1}^t$, $f_{x}^{t-1}$, and $f_{\alpha_3}^t$ are concatenated as inputs, producing an output split into $f_{u_s}^t$, $f_r^t$ and $f_{v_s}^t$, where $f_r^t$ denotes the residual feature.

\textit{2) Channel Fusion Module.} As illustrated in Fig.~\ref{fig:moduleA} (b), the U-CFM$(\cdot)$ and V-CFM$(\cdot)$ are channel attention-based Transformers~\cite{zamir2022restormer} that refine the interaction between $f_x^{t-1}$ and the spatially fused outputs $f_{u_s}^{t}$ or $f_{v_s}^{t}$ to update $f_{u}^t$ and $f_{v}^t$, respectively, while $f_x^{t-1}$ remains unchanged based on our formulation. 

\textbf{Update $f_{x}^t$}: 
As illustrated in Fig.~\ref{fig:moduleA} (c), for the optimization of variable $f_x^{t-1}$ from the previous stage, utilizing the updated results $f_{u}^t$ and $f_{v}^t$ from Eq.~\eqref{eq:channel-fusion-u} and Eq.~\eqref{eq:channel-fusion-v}, we proceed through the Data Consistency Module $\rm DCM(\cdot)$ to obtain $f_{x_p}^t$ according to Eq.~\eqref{eq:hqs-update-X-M}, as expressed in the following equation:
\begin{equation}
\small
f_{x_p}^t = {\rm DCM}(f_{u}^t, {f_{y_2}}, f_{v}^t),
\end{equation}
where $f_{x_p}^t$ refers to the preliminary update result of $f_x^{t-1}$. Then, we fuse optimized variables $f_{u}^{t}, f_{x_p}^{t}, f_{v}^{t}$ and do dimension reduction using $\rm MLP(\cdot)$ to further update $f_{x_p}^t$. Subsequently, we add the residual feature $f_r^t$ to obtain $f_x^t$. 
\begin{equation}
\small
f_x^t = {\rm MLP}([f_{u}^t,f_{x_p}^t,f_{v}^t]) + f_r^t.
\end{equation}

Finally, we transmit optimized variables, $f_{\alpha_1}^t$, $f_{x}^t$, and $f_{\alpha_3}^t$ to the next stage.

\vspace{1mm}
\textbf{3) HDR Image Reconstruction.} 
After completing all the unfolding reconstruction stages, we obtain the final reconstructed HDR feature $f_x^{\rm T}$. 
To ensure stability in HDR image reconstruction, we employ a residual strategy and then project the feature into the reconstructed HDR image $\hat{x}$. 
The reconstruction process is defined as follows:
\begin{equation} 
\small
\hat{x} =  {\rm Sigmoid} ( {\rm Conv} (f_x^{\rm T} + {\rm {Conv} }({f_{y_2}}) ) ),  \\
\label{eq:ImageRecon}
\end{equation}
where $\rm {Sigmoid} (\cdot)$ and $\rm {Conv} (\cdot)$ represent the Sigmoid activation function and convolutional operation, respectively.

\subsection{Training Loss}
Our model is trained end-to-end with the linear combination of $\mathcal{L}_{1}$ loss and perceptual loss $\mathcal{L}_{p}$. Considering that computing the loss in the HDR domain leads to less effective training~\cite{kalantari2017deep}, we calculate the loss in the tone-mapped domain by applying the $\mu$-law function, the total loss $\mathcal{L}$ is:
\begin{equation}
\small
\mathcal{L} = \| {\rm{\tau}}(x)-{\rm{\tau}}(\hat{x}) \|_1 + \eta \sum_k \| \phi_k ({\rm{\tau}}(x)) -  \phi_k ({\rm{\tau}}(\hat{x}))  \|_1,
\label{eq:l1 loss and perceptual loss}
\end{equation}
where ${\rm{\tau}}(x) = \frac{\text{log}(1 + \mu x)}{\text{log}(1 + \mu)}$ is the tone-mapping function with  $\mu$ = 5000. $\phi_k (\cdot)$ is the feature from the $k$-th layer of the VGG-19~\cite{simonyan2015very}, and $\eta=0.005$ is the weighting parameter.

\section{Experiments}

This section validates the performance of AFUNet through extensive quantitative and qualitative comparisons, along with ablation studies. 
\textit{Additional quantitative comparisons, qualitative results, and detailed ablation studies are included in the supplementary material.}

\subsection{Experimental Setups}

\noindent \textbf{Dataset.}
All methods are trained using three publicly available and widely used datasets, employing the same training settings: Kalantari’s dataset~\cite{kalantari2017deep}, which consists of 74 samples for training and 15 for testing, Tel's dataset~\cite{tel2023alignment}, which contains 108 training samples and 36 testing samples, Hu’s dataset~\cite{hu2013hdr} with 85 samples for training and 15 samples for testing.
Moreover, to further validate the model’s generalizability, we test on Tursun’s dataset~\cite{tursun2016objective} only for qualitative assessment, which lacks ground truth.

\vspace{1mm}
\noindent \textbf{Evaluation Metrics.}
We use peak signal-to-noise ratio (PSNR) and SSIM~\cite{wang2004image} as evaluation metrics, calculating both metrics in the linear and tone-mapped domains, denoted as ‘-$l$’ and ‘-$\mu$’, respectively. 
Moreover, we adopt HDR-VDP2~\cite{mantiuk2011hdr} that measures the human visual difference between results and targets. 

\vspace{1mm}
\noindent \textbf{Implementation Details.}
Our implementation is in PyTorch, and the AFUNet model is configured with a default of 4 stages. Each stage is comprised of 2 SAMs, 1 SFM, and 2 CFMs.
During training, we sample \(128 \times 128\) patches from the dataset and apply data augmentation techniques including random cropping, rotation, and flipping.
We use the Adam optimizer with a batch size of 6 and an initial learning rate of \(5 \times 10^{-4}\), which is decayed to \(5 \times 10^{-6}\) using cosine decay.
The model is trained for 400 epochs on a single NVIDIA GeForce 4090 GPU.

\begin{table}[t]
\small
\setlength{\tabcolsep}{3.0pt}
\centering
\begin{tabular}{c|ccccc}
\toprule
Method          & PSNR-$\mu$ & PSNR-$l$ & SSIM-$\mu$ & SSIM-$l$ & HDR-VDP2 \\
\midrule
DHDR & 41.64 & 40.91 & 0.9869 & 0.9858 & 60.50 \\
AHDR & 43.62 & 41.03 & 0.9900 & 0.9862 & 62.30 \\
NHDRR & 42.41 & 41.08 & 0.9887 & 0.9861 & 61.21 \\
HDR-GAN & 43.92 & 41.57 & 0.9905 & 0.9865 & 65.45 \\
ADNet & 44.37 & 41.88 & 0.9917 & 0.9892 & 66.02 \\
APNT & 43.94 & 41.61 & 0.9898 & 0.9879 & 64.05 \\
FlexHDR & 44.35 & \cellcolor{orange!30}42.60 & \cellcolor{red!30}0.9931 & \cellcolor{orange!30}0.9902 & 66.56 \\
CA-ViT & 44.32 & 42.18 & 0.9916 & 0.9884 & 66.03 \\
HyHDR & 44.64 & 42.47 & 0.9915 & 0.9894 & 66.05 \\
DiffHDR & 44.11 & 41.73 & 0.9911 & 0.9885 & 65.52 \\
SCTNet & 44.43 & 42.21 & 0.9918 & 0.9891 & \cellcolor{yellow!30}66.64 \\
LFDiff & \cellcolor{orange!30}44.76 & \cellcolor{yellow!30}42.59 & \cellcolor{yellow!30}0.9919 & \cellcolor{red!30}0.9906 & 66.54 \\
SAFNet & \cellcolor{yellow!30}44.66 & \cellcolor{red!30}43.18 & \cellcolor{yellow!30}0.9919 & \cellcolor{yellow!30}0.9901 & \cellcolor{orange!30}66.69 \\
RFG-HDR & 44.21 & 42.16 & 0.9915 & 0.9893 & 66.47 \\
Ours & \cellcolor{red!30}44.91 & \cellcolor{yellow!30}42.59 & \cellcolor{orange!30}0.9923 & \cellcolor{red!30}0.9906 & \cellcolor{red!30}66.75 \\
\bottomrule
\end{tabular}
\vspace{-6pt}
\caption{
Quantitative comparisons on Kalantari's dataset~\cite{kalantari2017deep}. The top three performances are highlighted in red, orange, and yellow backgrounds, respectively.
}
\vspace{-6pt}
\label{tab: Kal}
\end{table}

\begin{figure*}
\centering
\includegraphics[width=1.0\textwidth]{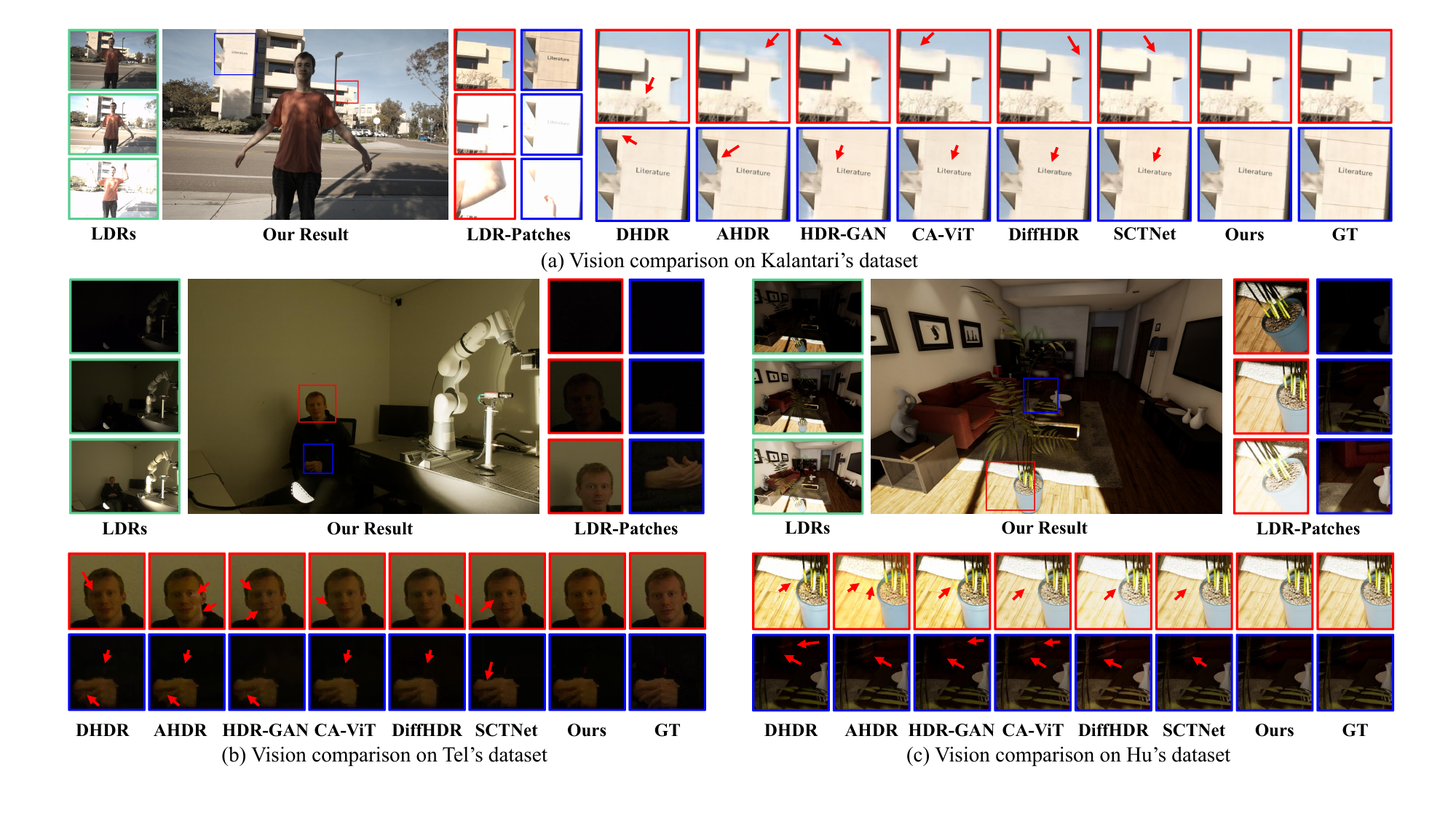}
\vspace{-18pt}
\caption{
Qualitative comparison between our method and state-of-the-art methods on three datasets: (a) Kalantari's dataset~\cite{kalantari2017deep}, (b) Tel's dataset~\cite{tel2023alignment}, and (c) Hu's dataset~\cite{hu2013hdr}. 
}
\vspace{-12pt}
\label{fig:exhib}
\end{figure*}

\subsection{Comparison with the State-of-the-art Methods}
To comprehensively evaluate our model, we compare it with several conventional and state-of-the-art methods from various categories. These include CNN-based methods, \textit{i.e.},~DHDR~\cite{wu2018deep}, AHDR~\cite{yan2019attention}, NHDRR~\cite{yan2020deep}, ADNet~\cite{liu2021adnet}, APNT~\cite{chen2022attention}, FlexHDR~\cite{catley2022flexhdr} and SAFNet~\cite{kong2024safnet};
Generative Adversarial Network (GAN)-based methods, \textit{i.e.},~HDR-GAN~\cite{niu2021hdr}; 
Transformer-based methods, \textit{i.e.},~CA-ViT~\cite{liu2022ghost}, HyHDR~\cite{yan2023unified},  SCTNet~\cite{tel2023alignment} and RFG-HDR~\cite{lee2024rfg}; 
Diffusion model-based methods, \textit{i.e.},~DiffHDR~\cite{yan2023towards} and LFDiff~\cite{hu2024generating}.

\begin{table}[t]
\small
\setlength{\tabcolsep}{3.0pt}
\centering
\begin{tabular}{c|ccccc}
\toprule
Method          & PSNR-$\mu$ & PSNR-$l$ & SSIM-$\mu$ & SSIM-$l$ & HDR-VDP2 \\
\midrule
DHDR & 41.13 & 41.20 & 0.9870 & 0.9941 & 70.82 \\
AHDR & 45.76 & 49.22 & 0.9956 & 0.9980 & 75.04 \\
NHDRR & 45.15 & 48.75 & 0.9956 & 0.9981 & 74.86 \\
HDR-GAN & 45.86 & 49.14 & 0.9945 & \cellcolor{yellow!30}0.9989 & 75.19 \\
APNT & 46.41 & 47.97 & 0.9953 & 0.9986 & 73.06 \\
CA-ViT & 48.10 & 51.17 & 0.9947 & \cellcolor{yellow!30}0.9989 & 77.12 \\
HyHDR & \cellcolor{yellow!30}48.46 & \cellcolor{yellow!30}51.91 & \cellcolor{yellow!30}0.9959 & \cellcolor{orange!30}0.9991 & \cellcolor{yellow!30}77.24 \\
DiffHDR & 48.03 & 50.23 & 0.9954 & \cellcolor{yellow!30}0.9989 & 76.22 \\
SCTNet & 48.10 & 51.14 & \cellcolor{orange!30}0.9963 & \cellcolor{orange!30}0.9991 & 77.14 \\
LFDiff & \cellcolor{orange!30}48.74 & \cellcolor{orange!30}52.10 & \cellcolor{red!30}0.9968 & \cellcolor{red!30}0.9993 & \cellcolor{orange!30}77.35 \\
Ours & \cellcolor{red!30}48.83 & \cellcolor{red!30}52.13 & \cellcolor{red!30}0.9968 & \cellcolor{orange!30}0.9991 & \cellcolor{red!30}77.44 \\
\bottomrule
\end{tabular}
\vspace{-6pt}
\caption{
Quantitative comparisons on Hu's dataset~\cite{hu2013hdr}. 
}
\vspace{-10pt}
\label{tab: Hu}
\end{table}

\begin{table}[t]
\small
\setlength{\tabcolsep}{3.0pt}
\centering
\begin{tabular}{c|ccccc}
\toprule
Method          & PSNR-$\mu$ & PSNR-$l$ & SSIM-$\mu$ & SSIM-$l$ & HDR-VDP2 \\
\midrule
DHDR & 40.05 & 43.37 & 0.9794 & 0.9924 & 67.09 \\
AHDR & 42.08 & 45.30 & 0.9837 & 0.9943 & 68.80 \\
NHDRR & 36.68 & 39.61 & 0.9590 & 0.9853 & 65.41 \\
HDR-GAN & 41.71 & 44.87 & 0.9832 & \cellcolor{yellow!30}0.9949 & 69.57 \\
CA-ViT & \cellcolor{yellow!30}42.39 & \cellcolor{yellow!30}46.35 & \cellcolor{yellow!30}0.9844 & 0.9948 & 69.23 \\
DiffHDR & 42.18 & 45.63 & 0.9841 & 0.9946 & \cellcolor{yellow!30}69.88 \\
SCTNet & \cellcolor{orange!30}42.55 & \cellcolor{orange!30}47.51 & \cellcolor{orange!30}0.9850 & \cellcolor{orange!30}0.9952 & \cellcolor{orange!30}70.66 \\
Ours & \cellcolor{red!30}43.31 & \cellcolor{red!30}47.83 & \cellcolor{red!30}0.9876 & \cellcolor{red!30}0.9959 & \cellcolor{red!30}71.08 \\
\bottomrule
\end{tabular}
\vspace{-6pt}
\caption{
Quantitative comparisons on Tel's dataset~\cite{tel2023alignment}. 
}
\vspace{-14pt}
\label{tab: Tel}
\end{table}

\noindent \textbf{Qualitative Comparison.}
The quantitative results of AFUNet on three widely-used datasets,~\textit{i.e.} Kalantari’s dataset~\cite{kalantari2017deep}, Hu’s dataset~\cite{hu2013hdr}, and Tel's dataset~\cite{tel2023alignment}, are presented in Tab.~\ref{tab: Kal}, Tab.~\ref{tab: Hu} and Tab.~\ref{tab: Tel}, respectively.
Our method is compared with classical and state-of-the-art approaches, which include challenging scenarios such as under/over-exposure regions and large misalignment, which are prone to causing ghosting artifacts. 
Notably, AFUNet exhibits great improvement over previous methods, surpassing Transformer-based methods CA-ViT~\cite{liu2022ghost} and SCTNet~\cite{tel2023alignment} by 0.59 dB and 0.48 dB in PSNR-$\mu$, 0.41 dB and 0.38 dB in PSNR-$l$, respectively, on Kalantari’s dataset. 
In addition, AFUNet outperforms the other leading methods, achieving better reconstruction performance. Compared to LFDiff~\cite{hu2024generating}, AFUNet demonstrates improvements of 0.14 dB and 0.04 dB in PSNR-$\mu$ and SSIM-$\mu$ on Kalantari’s dataset. Furthermore, it surpasses SAFNet~\cite{kong2024safnet} by 0.25 dB in PSNR-$\mu$ and 0.04 dB in SSIM-$\mu$ on Kalantari’s dataset, which represent greater reconstruction performance. 

\begin{figure}[t]
\centering
\includegraphics[width=1.0\linewidth]{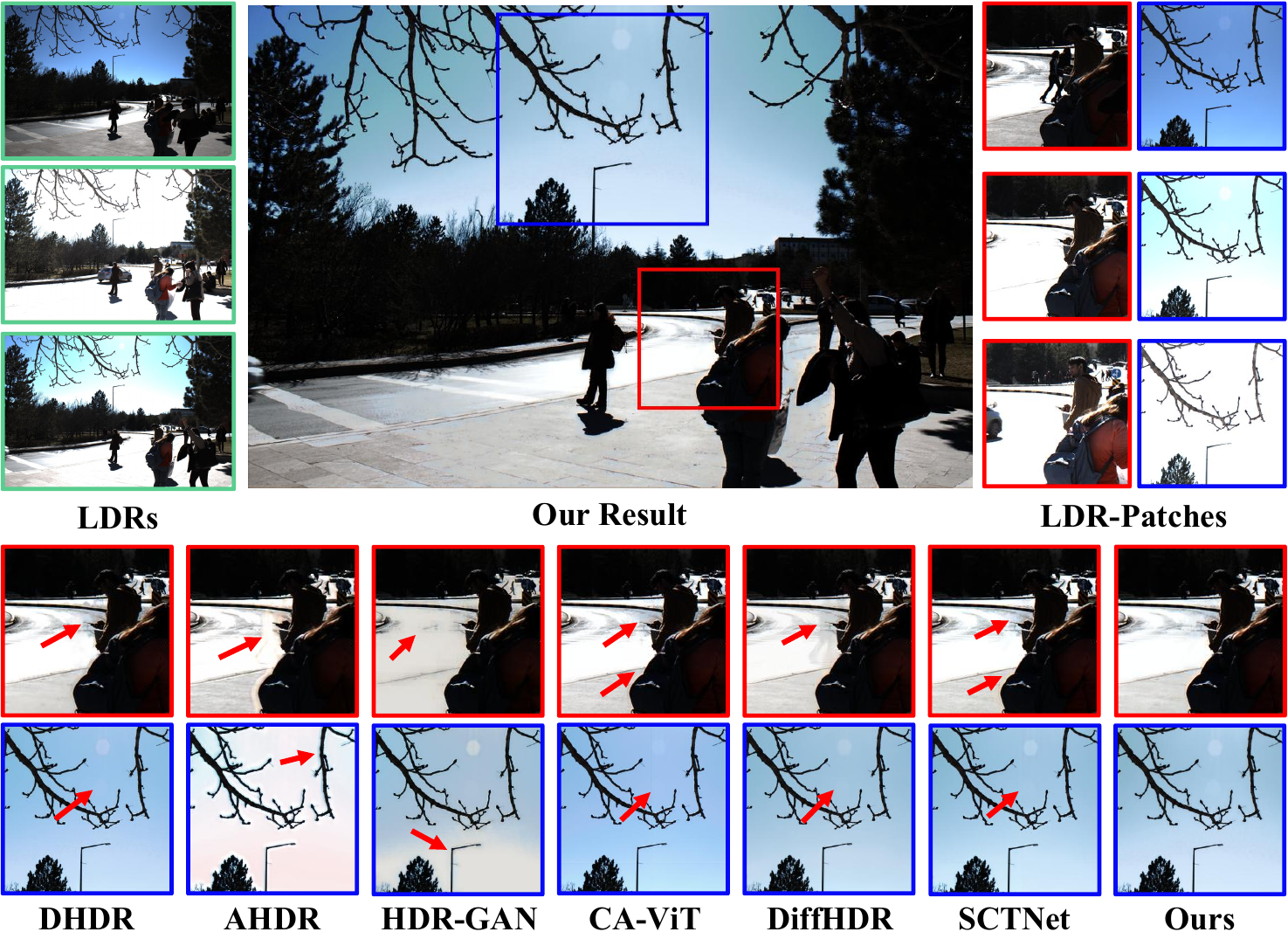}
\vspace{-12pt}
\caption{
Qualitative comparison between our method and state-of-the-art methods on Tursen's dataset~\cite{tursun2016objective} without ground truth.
}
\vspace{-6pt}
\label{fig:Tur_visual}
\end{figure}

\noindent \textbf{Quantitative Comparison.}
The visual comparisons on Kalantari’s dataset~\cite{kalantari2017deep}, Tel’s dataset~\cite{tel2023alignment} and Hu's dataset~\cite{hu2013hdr} are shown in Fig.~\ref{fig:exhib}.
We can observe that our proposed AFUNet achieves more complete scene reconstruction and retains more details. 
AFUNet demonstrates strong reconstruction capabilities for challenging patches,
with a substantial reduction in ghosting artifacts.
To assess the generalization capability of the proposed HDR imaging method, we evaluate the model trained on Kalantari’s dataset~\cite{kalantari2017deep} and tested on Tursun’s dataset~\cite{tursun2016objective}, which lacks ground truth. 
Therefore, we only use our human subjective perception to judge the model's performance in this comparison. The visual comparison on Tursun's dataset~\cite{tursun2016objective} is shown in Fig.~\ref{fig:Tur_visual}.
We attribute the strong performance to our carefully designed HDR optimization reconstruction algorithm and the unfolding framework. The alignment and fusion sub-problems can complement each other during the reconstruction process.
Moreover, cross-iterative alignment and fusion synergy significantly leads to more effective optimization, allowing us to generate high-quality HDR images with improved perceptual quality. 

\subsection{Ablation Study}
To investigate the effectiveness of each key component and the unfolding paradigm for HDR reconstruction, we conducted thorough ablation studies on Kalantari's dataset~\cite{kalantari2017deep} using the following variants of our model: 
(1) \textbf{M1}: SFM.
(2) \textbf{M2}: Adding SAM into M1. 
(3) \textbf{M3}: Adding CFM into M1. 
(4) \textbf{M4}: Adding DCM into M1.

\begin{table}[t]
\footnotesize
\setlength{\tabcolsep}{2.0pt}
\centering
\begin{tabular}{c|cccc|cccc}
\toprule
Models          & SFM & SAM & CFM & DCM & PSNR-$\mu$ & PSNR-$l$ & SSIM-$\mu$ & SSIM-$l$\\
\midrule
M1 &\checkmark& & & & 43.94 & 42.04 & 0.9917 & 0.9890 \\
M2 &\checkmark&\checkmark& & & 44.48 & 42.40 & 0.9918 & 0.9893 \\
M3 &\checkmark& &\checkmark & &44.62 &42.56 &0.9921 &0.9895 \\
M4 &\checkmark& & &\checkmark &44.45 &42.39& 0.9919& 0.9899 \\
\midrule
AFUNet &\checkmark&\checkmark&\checkmark&\checkmark& 44.91 & 42.59 & 0.9923 & 0.9906 \\
\bottomrule
\end{tabular}
\vspace{-6pt}
\caption{
Ablation study of different components in the proposed unfolding framework on Kalantari’s dataset~\cite{kalantari2017deep}. 
}
\vspace{-10pt}
\label{tab:ablation model cases}
\end{table}

\noindent \textbf{Different Components.}
We conduct ablation experiments to evaluate the contribution of each component within  AFUNet.
The experimental cases and corresponding qualitative results are presented in Tab.~\ref{tab:ablation model cases}. 
The results demonstrate the effectiveness of the SAM, CFM, and DCM, with each component contributing to notable improvements in the performance of our method. 
Specifically, our thorough ablation studies conclusively show the effectiveness and necessity of the alignment module in the M2 cases, showing that incorporating the alignment process into the fusion process can improve the final reconstruction quality.

\noindent \textbf{Different HDR Reconstruction Paradigms.}
We investigate the effectiveness of our proposed progressive alignment and fusion unfolding paradigm, introducing a novel perspective on analyzing our framework. 
Each stage of our model consists of two processes: alignment and fusion. As shown in Tab.~\ref{tab:ablation different paradigm}, we can reconstruct the framework into two distinct paradigms.
Specifically, we explore the cross-iterative Alignment-Fusion paradigm denoted as ``AF", and the cross-iterative Fusion-Alignment paradigm denoted as ``FA" as alternative perspectives for evaluating our unfolding algorithm. As shown in Tab.~\ref{tab:ablation different paradigm}, the ``FA” paradigm demonstrates a slight performance decline compared to our ``AF” paradigm, highlighting the effectiveness of the alignment process while maintaining competitive quantitative results due to our robust formulation.

\begin{table}[t]
\small
\setlength{\tabcolsep}{4.0pt}
\centering
\begin{tabular}{c|cc|cccc}
\toprule
Models          & AF & FA & PSNR-$\mu$ & PSNR-$l$ & SSIM-$\mu$ & SSIM-$l$\\
\midrule
P1 &\checkmark&          & 44.91 & 42.59 & 0.9923& 0.9906\\
P2 &          &\checkmark&44.72 &42.32 &0.9923 &0.9904  \\
\bottomrule
\end{tabular}
\vspace{-6pt}
\caption{
Ablation study of different paradigms. ``AF" first conducts alignment followed by fusion, and ``FA" first conducts fusion followed by alignment, both in an alternating sequence.
}
\label{tab:ablation different paradigm}
\end{table}

\begin{table}[t]
\small
\setlength{\tabcolsep}{8.0pt}
\centering
\begin{tabular}{c|cccc}
\toprule
Stages    & PSNR-$\mu$ & PSNR-$l$ & SSIM-$\mu$ & SSIM-$l$  \\
\midrule
2 & 44.40 & 41.45 & 0.9918 & 0.9881  \\
3 & 44.83 & 42.62 & 0.9923 & 0.9903  \\
4 & 44.91 & 42.59 & 0.9923 & 0.9906  \\
5 & 44.85 & 42.91 & 0.9923 & 0.9908  \\
6 & 44.93 & 42.84 & 0.9923 & 0.9910  \\
\bottomrule
\end{tabular}
\vspace{-6pt}
\caption{
The impact of different numbers of iterative reconstruction stages in AFUNet on Kalantari’s dataset~\cite{kalantari2017deep}.
}
\label{tab:stages num exp results}
\vspace{-10pt}
\end{table}

\noindent \textbf{Number of Stages.}
We investigate the impact of different numbers of unfolding iterative stages in AFUNet, specifically 2, 3, 4 (default), 5, and 6, to explore their influence on model performance. 
As shown in Tab.~\ref{tab:stages num exp results}, there is a correlation between the number of stages and reconstruction performance, demonstrating the effectiveness of our iterative design. Fewer stages result in lower performance compared to the default configuration but the 3-stage setting surpasses the previous stage-of-the-art method in PSNR-$\mu$ and SSIM-$\mu$, showcasing the superior reconstruction ability of our AFUNet. 
While using 5 and 6 stages yields slight performance improvements over 4 stages, it comes at the cost of increased training time and model complexity. 
Thus, we select 4 stages as the default setting, striking an optimal balance between performance and model complexity.

\section{Conclusion}

In this paper, we propose AFUNet, a novel and effective cross-iterative Alignment and Fusion deep Unfolding Network for HDR reconstruction. 
We first formulate the HDR reconstruction objective, introducing spatial correspondence priors among LDR images. 
Then, we derive the HDR reconstruction process in detail, which is subsequently unfolded into an end-to-end trainable network. 
Each iteration consists of two sub-problems—alignment and fusion—where we carefully design corresponding modules to iteratively optimize the overall problem. 
Extensive experiments show that AFUNet excels in producing realistic HDR images with more detail and less ghosting, outperforming state-of-the-art methods.  

\textbf{Acknowledgments}. This work was supported in part by the National Natural Science Foundation of China under Grant 62201387 and in part by the Fundamental Research Funds for the Central Universities.

{
    \small
    \bibliographystyle{ieeenat_fullname}
    \bibliography{main}
}

\end{document}